# HOW MUCH CAN GALACTIC ABERRATION IMPACT THE LINK BETWEEN RADIO (ICRF) AND OPTICAL (GCRF) REFERENCE FRAMES*

Zinovy Malkin[1,2]

[1]Pulkovo Observatory, St. Petersburg, Russia
[2]St. Petersburg State University, St. Petersburg, Russia

**Abstract.** *The highly-accurate optical reference frame GCRF (Gaia Celestial Reference Frame) is expected to be available in several years. By the same time, a new version of radio reference frame ICRF (International Celestial Reference Frame) will be also published. The link of GCRF to ICRF will be defined by means of computation of the orientation angles between the two frames using the common extragalactic objects observed in both radio (VLBI) and optics (Gaia). Taking into account the expected accuracy of ICRF and GCRF of the first tens microarcseconds, the link between them should be defined at a microarcsecond level, which requires using the most accurate algorithms and models. One of such models is the Galactic aberration in proper motions, which is not included in the data processing yet. In this paper its impact on the ICRF-Gaia link is estimated. Preliminary results showed that this impact is at a level of about 1 microarcsecond.*

## Introduction

Gaia mission was successfully started in December 2013 and its main scientific program began in July 2014. One of the main result of the Gaia mission will be a new highly-accurate celestial reference frame GCRF (Gaia Celestial Reference Frame) at a level of a few tens microarcseconds. The final GCRF version is expected to be available in early 2020s. On the other hand, a new version of the VLBI-based celestial reference frame of similar accuracy ICRF3 (International Celestial Reference Frame, 3rd realization) is planned to be published by the end of this decade (Jacobs et al., 2013).

Both frames, ICRF and GCRF, should realize the ICRS (International Celestial Reference System) as accurately as possible. Currently the second ICRF realization, ICRF2, is the official IAU-recommended ICRS realization. It can be supposed that the ICRF3 catalog will be linked to the ICRF2 catalog in a way similar to that used for linking ICRF2 to the first ICRF (Fey et al., 2015). Then the GCRF will be linked to ICRF3. The link is, in fact, determination of the orientation angles between two catalogs using a list of common extragalactic objects observed in both radio (VLBI) and optics (Gaia). Such a link should be performed with a microarcsecond accuracy to not compromise the high internal accuracy of the Gaia frame. To achieve this, in particular, the most precise algorithms and models should be used for data processing. One of such models is the Galactic aberration in proper motion (GA). This work is devoted to preliminary estimate the GA impact on the ICRF–GCRF link.

## Impact of the GA on the orientation angles between ICRF and GCRF

To estimate the impact of the GA on the orientation angles between ICRF and GCRF the following numerical test was performed. First, possible link sources were selected from the ICRF2 catalog using source type and optical magnitude as criteria of selection. The final link source list comprises of 688 sources of AGN type and with visual magnitude 18$^m$ or brighter, following Bourda et al. (2008). The latest version of the OCARS catalog[1] (Malkin & Titov, 2008) was used to get the source type and magnitude. Distribution of the sources over the sky is depicted in Fig. 1. One can see that the distribution of the link sources over the sky is near uniform with some exhaustion in the Galactic equatorial zone.

---

[1] http://www.gao.spb.ru/english/as/ac_vlbi/ocars.txt

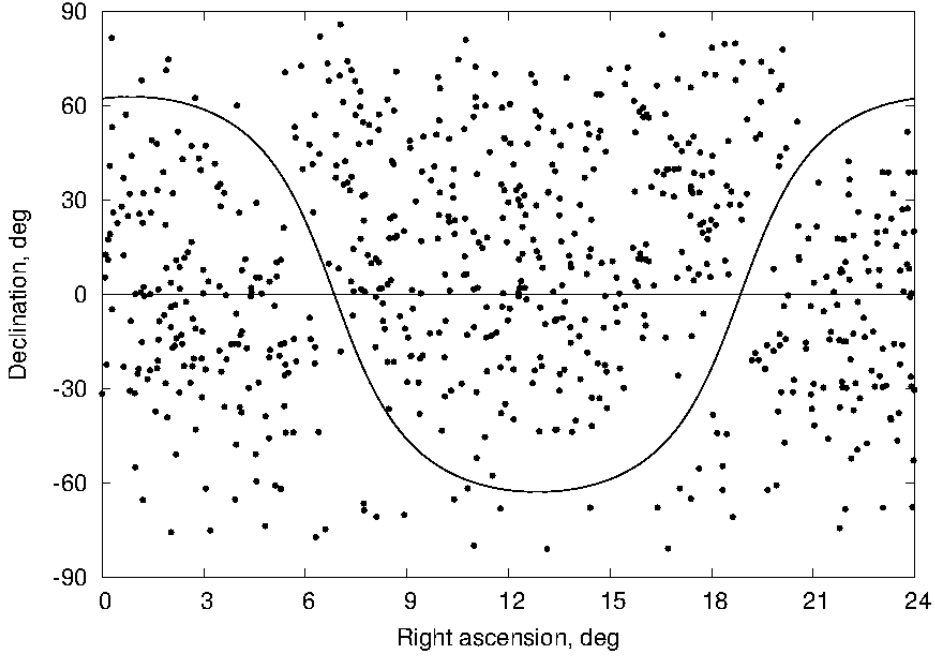

**Fig. 1.** Distribution of 688 ICRF2–GCRF link sources over the sky. Points correspond to the source position, and solid line corresponds to the Galactic equators.

Then, it looks reasonable to choose the epoch $t_0 = 2017.0$ as the epoch of comparison of Gaia and ICRF catalogs. It is the expected middle epoch of the 5-year Gaia observation period started in July 2014. As to Gaia data, it can be expected that the GCRF catalog will be brought to the epoch $t_0$ by the usual astrometric reduction procedures using the object proper motions deduced from the same Gaia observations. As to ICRF catalog, the latter does not contain radio source proper motions, the catalog is considered as epochless. The GA is the only known and precisely modelled effect that causes the ICRF sources proper motion. So, to bring the selected link sources to the epoch $t_0$ the following equations were used:

$$\alpha(t_0) = \alpha(t_i) - \mu_\alpha(t_i - t_0), \\ \delta(t_0) = \delta(t_i) - \mu_\delta(t_i - t_0), \quad (1)$$

where $t_i$ is the mean epoch of observations of $i$-th radio source in ICRF2, $\alpha(t_i)$ and $\delta(t_i)$ are the coordinates of $i$-th radio source in ICRF2 at epoch $t_i$, $\mu_\alpha$ and $\mu_\delta$ is the GA-derived proper motion of $i$-th radio source in right ascension and declination, respectively, computed according to Malkin (2014).

In such a way, two catalogs were obtained. The first one is merely a selection of 688 sources from ICRF2. Such a catalog can be used for the ICRF–GCRF link if the ICRF is considered as epochless, which is currently the case. The second catalog contains positions of the same sources transferred to the epoch $t_0$ for the GA-induced proper motions. Mutual orientation between these two catalogs is defined by three angles $A_1$, $A_2$, and $A_3$ according to classical equations:

$$\Delta\alpha = A_1 \cos\alpha \tan\delta + A_2 \sin\alpha \tan\delta - A_3, \\ \Delta\delta = -A_1 \sin\alpha \quad + A_2 \cos\alpha . \quad (2)$$

Result of computation is presented in Table 1. The orientation angles given in the Table are, in fact, the difference in the orientation angles between ICRF and GCRF computed with and without taking the GA effect into account, supposing the Gaia source positions are given at the mean Gaia epoch $t_0 = 2017.0$.

**Table 1.** Orientation angles between original catalog (688 optically bright AGNs with the ICRF2 positions) and catalog converted to the epoch 2017.0 for the Galactic aberration, μas.

| $A_1$ | $A_2$ | $A_3$ |
|---|---|---|
| 1.32 ± 0.50 | –0.06 ± 0.49 | 0.28 ± 0.42 |

**Conclusion**

A realistic numerical test was performed to estimate the impact of the GA on the orientation angles between ICRF and GCRF. The result of this test has shown that this impact is practically negligible, its value is at the level of 1 μas. Such a small effect can be explained by sufficiently uniform distribution of the link sources over the sky (see Fig. 1). However, it is not simple now to foreseen the exact list of objects that will be actually used for the ICRF–GCRF link, and how good they will be distributed over the sky. It seems to be necessary to repeat such a test later using a source list more close to actual one. Indeed, the best solution of this problem will be implementation of the GA in the routine practice of the data processing, e.g., using the strict method descripted by Malkin (2014). It should be kept in mind that such a change in the data processing algorithm should be made simultaneously in all analysis centers. Otherwise, the results obtained at different centers will be hardly compatible to each other.